\documentclass[twocolumn,showpacs,10pt,superscriptaddress]{revtex4}
\usepackage{bbm}
\usepackage{amsmath}
\usepackage{amssymb}
\usepackage{txfonts}
\usepackage[dvips]{graphicx}
\usepackage{dcolumn}
\usepackage{array}
\usepackage{color, soul}
\newcommand{\upcite}[1]{\textsuperscript{\textsuperscript{\cite{#1}}}}

\begin{document}

\preprint{}

\title{Topological phase transition based on the attractive Hubbard model}

\author{Fenghua Qi}
\affiliation{Center for Quantum Transport and Thermal Energy Science, School of Physics and Technology, Nanjing Normal University, Nanjing, 210023, China}
\affiliation{School of Electronic Engineering, Nanjing Xiaozhuang University, Nanjing 211171, China}
\author{Jun Cao}
\affiliation{Center for Quantum Transport and Thermal Energy Science, School of Physics and Technology, Nanjing Normal University, Nanjing, 210023, China}
\affiliation{School of Electronic Engineering, Nanjing Xiaozhuang University, Nanjing 211171, China}
\author{Jie Cao}
\affiliation{College of Science, Hohai University, Nanjing 210098, China}
\author{Xiao Li\footnote{Authors to whom correspondence should be addressed. Electronic mails: lixiao@njnu.edu.cn}}
\affiliation{Center for Quantum Transport and Thermal Energy Science, School of Physics and Technology, Nanjing Normal University, Nanjing, 210023, China}
\author{Lifa Zhang\footnote{Authors to whom correspondence should be addressed. Electronic mails: phyzlf@njnu.edu.cn}}
\affiliation{Center for Quantum Transport and Thermal Energy Science, School of Physics and Technology, Nanjing Normal University, Nanjing, 210023, China}

\date{\today}

\begin{abstract}
We theoretically investigate the effect of an attractive on-site interaction on the two-band magnetic Dirac fermion model based on a square lattice system. When the attractive fermion interaction is taken into account
by the mean-field approximation, a phase diagram is obtained. It is found that a quantum phase transition from
a band insulator state to quantum anomalous Hall state occurs with increased attractive interaction. For an existing quantum anomalous Hall state, the attractive interaction enlarges its nontrivial
band gap and makes the topological edge states more localized, which protects the transport of linear-dispersive edge states
against finite-size and further disorder effects.
\end{abstract}


\maketitle

\section{\label{sec1}Introduction}
For the past decade, the subject of topological insulating states has attracted great attention in condensed-matter physics.\upcite{Kane, Bernevig, Fu.L, Murakami, Moore2, Qi.X.L., Dai.X., Hsieh, Haijun, Chen, Feng, Emilio}
Topological insulating states have insulating energy gaps in the bulk but gapless boundary states as a defining feature. Quantum spin Hall (QSH)\upcite{Haldane, Qi.X.L.2} and quantum anomalous Hall (QAH)\upcite{Rui Yu, Cui-Zu Chang, Jing Wang} states are two typical examples of two-dimensional topological insulating states. The QAH state, with time-reversal symmetry breaking, harbors chiral edge states that are robust against disorder, which is similar to quantum Hall effect but does not need external magnetic field. The QSH state, usually regarded as two copies of QAH state, harbors helical edge states with the time-reversal symmetry protection.\upcite{Qi.X.L.1, M. Z. Hasan} These topological edge states of the topological insulating states provide dissipationless quantum transport, which have enormous potential application in low-power spintronics and quantum computation.\upcite{Fabian, Moore}


The band order of the low energy electronic structure of the topological insulating state determines the band topology. A band inversion process between low-energy states with distinct symmetries usually helps create a topological insulating state. For example, a QSH state can be obtained by inverting two parity-distinct bands of a band insulator (BI) with the inversion symmetry. The band inversion is conventionally
driven by a considerable spin-orbit coupling and/or applied strain field. With a repulsive electron-electron interaction considered, the band inversion can occur, leading to a topological Mott insulator.\upcite{XZhang} Besides, new mechanisms of the band inversion processes are still lacking and highly desirable to achieve topological insulating states.

While a lot of works have discussed the effect of the repulsive fermion interaction on the band
inversion,\upcite{XZhang, Raghu, LWang, Yoshida, JCao} little attention has been drawn to the attractive electron-electron interaction, which may play a role in the band inversion and associated topological phase transition. The attractive interaction mostly could be realized in cold atoms loaded into optical lattice systems and be tuned by magnetic-field Feshbach resonances or changing the optical lattice depth.\upcite{Timmermans, Regal, Strohmaier} Its roles in the superconductivity have been fully addressed.\upcite{Q. J. Chen, Stefano Giorgini, Chih-Chun Chien, Roscilde, Fialko, Poletti, Zurn, Ruhman} Here, we consider the attractive interaction as a new mechanism of the band inversion process and then discuss the topological phase transition induced by it.

In this paper, we study the two-band magnetic Dirac fermion model\upcite{X.L. Qi, X.L. Qi2} based on a square lattice system. An attractive interaction is introduced between the on-site fermions with opposite spins, within the mean-field approximation. By the self-consistent calculation, it is found the attractive on-site interaction can induce a topological phase transition from the BI to the QAH state. The topological gap of the QAH state can be further enlarged and corresponding edge states become more localized, with the increase of the attractive interaction. The large bulk band gap and perfect Dirac-type dispersive edge states make the topological non-trivial state more robust against finite-size effects and further disorder effects to realize the energy-efficient edge transport. It is noted that  the two-band magnetic Dirac fermion model here is only an initial example to demonstrate the role of the attractive interaction and it is expected to broaden the scope of the topological phase transition. The same can apply to the creation of another topological insulating state, e.g., a QSH state, when the time-reversal counterpart of the magnetic Dirac model is also taken into account.


The rest of this paper is organized as follows. In Sec.~II, we introduce the two-band magnetic Dirac fermion model based on a square lattice system and the method of considering the attractive fermion interaction within the mean-field approximation. In Sec.~III, by the self-consistent calculation, we numerically present the influences of the attractive interaction on the phase transition, the topological energy gap and the edge states. Our conclusions are summarized in Sec.~IV.

\section{\label{sec2}MODEL AND METHODS}
We consider a low-energy magnetic Dirac fermion model in the simplest form,\upcite{X.L. Qi2}
\begin{equation}\label{little_hp}
h(\mathbf{k})=A(k_x\sigma_x+k_y\sigma_y)+m(k)\sigma_z.
\end{equation}
Here, the first term describes a massless Dirac state showing the momentum-dependent spin texture, while the second term gives a mass of $m(k)=m+B k^2$ to Dirac state with $m$ arising from the magnetization. $\sigma_{x,y,z}$ denote the spin Pauli matrices and $k$ is the wave vector. $A$, $B$ and $m$ are material-specific parameters. Throughout the paper, we use $A$ as the energy scale and set $A=1$. The values of $B$ and $m$ are crucial for determining the band order and corresponding band topology. If $mB<0$, this Hamiltonian describes a QAH insulator state with the Chern number = $\pm$1, otherwise if $mB>0$, this Hamiltonian describes a trivial magnetic band insulator with a vanishing Chern number.

The Hamiltonian (\ref{little_hp}) therefore embodies the topological phase transition  from the magnetic band insulator to the QAH state, which can be clearly expressed by a band inversion process. For convenience, we adopt the parameter $B<0$. As schematically illustrated in Fig.~1(a), we assume that the two-band system is originally in the topologically trivial BI phase with $m<0$. Considering the vertical component of the spin, the spin-up (red) and spin-down (black) subbands are well separated. When a special mechanism acting on the spin, e.g., an exchange field, is introduced, the spin-down subband is pushed down and the spin-up subband is lifted up. Hence, the band gap of $2|m|$ at $\mathbf k$=0 gets smaller, with an increased $m$. The subbands get closer and closer, and eventually experience a band inversion in Fig.~1(b), associated with a sign reversal of $m$ (i.e. $mB<0$). The spin-orbit coupling, embodied in the first term of the Hamiltonian (\ref{little_hp}), opens the gaps at the crossing points of the subbands, leading to the QAH state [Fig.~1(c)].

The key ingredient of the band inversion is to search for a mechanism to pull down the conduction band or push up the valence band. We will try to apply an attractive on-site interaction to realize the goal. In order to readily compute the on-site occupation number in attractive Hubbard model, we transform the low energy Hamiltonian (\ref{little_hp}) into a square lattice model, schematically shown in Fig.~1(d), by the simple substitutions, $k_{x,y}\rightarrow \sin(k_{x,y})$ and $k^2\rightarrow 4-2\cos(k_x)-2\cos(k_y)$ and the Fourier transformation from the momentum space to the real space. The lattice model immediately follows,
\begin{eqnarray}
H_S&=&\sum_{i,\sigma}smc_{i\sigma}^{\dagger}c_{i\sigma}-\sum_{\langle i,j\rangle,\sigma}sBc_{i\sigma}^{\dagger}c_{j\sigma}\notag\\&+&\frac{\textrm{i}A}{2}\sum_i[(-\nu c_{i\uparrow}^{\dagger}c_{i+1,\downarrow}+\nu c_{i\uparrow}^{\dagger}c_{i-1,\downarrow})+h.c.],
\end{eqnarray}
where $c_{i\sigma}^{\dagger}$ ($c_{i\sigma}$) is the creation (annihilation) operator of a fermion on site $i$ with the spin $\sigma$ and the lattice constant is set to unit. The first term is on-site exchange interaction, with $s=\pm 1$ for opposite spins. The second and third terms denote the nearest-neighbor hoppings between the same spin and opposite spins, respectively. $\langle i,j\rangle$ denotes the nearest neighboring pair of lattice sites. $\nu=1$ and $i$ correspond to the hopping along the $x$ and $y$ direction, respectively.
The lattice model including the fermion interaction is then obtained as,
\begin{equation}\label{HU}
H_U^0=H_S - 2U \sum_i n_{i\uparrow} n_{i\downarrow} - \mu \hat{N},
\end{equation}
where $U>0$ is the on-site attractive coupling parameter, the occupation number operator $n_{i\sigma}=c^\dagger_{i\sigma}c_{i\sigma}$ and the total number operator $\hat{N} = \sum_{i\sigma} n_{i\sigma}$. $\mu$ is the chemical potential. In the half-filling case, due to the fact that the density of state at the Fermi energy is exactly zero, the superconductivity cannot be induced in this attractive Hubbard model. Therefore, we will only consider the modulation of the band structures caused by the attractive interaction.

\begin{figure}
\includegraphics[width=7.5cm]{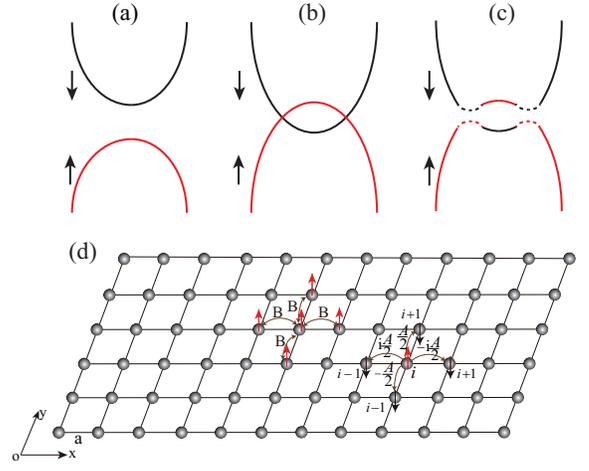}\caption{(Color online) (a)-(c) The evolution of the subband structures during the band inversion process. The red/black lines denote the spin-up/spin-down bands. The subbands initially separate, then go through each other and finally reopen an inverted gap, leading a topological phase transition from the BI to QAH state. (d) Schematic of a square lattice model. $A$ and $B$ are respectively the nearest-neighbor hopping energies between opposite spins and the same spin.}
\end{figure}

Using the mean-field approximation, we rewrite the Hamiltonian (\ref{HU}) as
\begin{equation}
H_U=H_S - \sum_i  \bigg[ n_{i\uparrow} \bigg( U \langle n_{i\downarrow} \rangle  + \mu \bigg) +  n_{i\downarrow} \bigg( U \langle n_{i\uparrow} \rangle  + \mu \bigg) \bigg].
\label{HU2}
\end{equation}
Here, the half-filling case is considered. In order to fulfill the constraint that $\langle n_{i\uparrow} \rangle + \langle n_{i\downarrow} \rangle =1$ at the half-filling case, the chemical potential lies inside the band gap and it is convenient to set $\mu=-U/2$. The Hamiltonian (\ref{HU2}) is solved by the self-consistent iterative method with a convergence threshold of $10^{-6}$ on the variance of the occupation number expectation, $\delta\langle n_{i\sigma} \rangle$. Transforming back to the continuum limit, the form of the Hamiltonian (\ref{little_hp}) keeps invariant but $m$ is replaced by a renormalized mass $M=m-\Delta(U)$,
with $\Delta(U) = U \left( \langle n_{i\downarrow} \rangle - \frac{1}{2} \right) = U \left( \frac{1}{2} - \langle n_{i\uparrow} \rangle \right) $ related to the calculated spin density. In the following section, we reconsider the topological criteria after taking into account the attractive interaction.
\section{\label{sec3}RESULTS AND DISCUSSION}
\subsection{The topological phase transition}
\begin{figure}
\includegraphics[width=7cm]{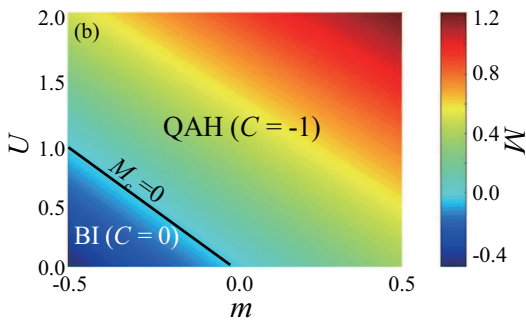} \caption{(Color online)  Phase diagram of the two-band QAH model with two variables $U$ and $m$. The color represents the value of the effective mass $M$. The black line is the phase boundary, which represents the critical effective mass $M_c=0$. Two different regions are respectively noted as BI and QAH.}
\end{figure}
Figure 2 demonstrates the calculated $M$ as a function of $U$ and $m$, based on the lattice model, where we adopt $B=-1$. It is seen that $M$ increases with $U$ ($m$), when $m$ ($U$) is kept invariant. A boundary line, corresponding to $M_c=0$, divides the phase diagram into two distinct regions, the  lower left side with $M<0$ and upper right side with $M>0$. It is also found that for a certain $m$ that is larger than zero, the sign of $M$ keeps unchanged with the increase of $U$. However, if $m<0$, $M$ can pass through the critical line $M_c=0$ with the increase of $U$. For the regions with $M\neq 0 $, there is a well-defined global gap, demonstrating an insulating state, which will be discussed in detail in the next subsection.

It is of interest to quantify the topological character of the phase diagram in Fig.~2, by calculating the Chern number of the insulating state,\upcite{Hasan}
\begin{equation}
C=\frac{1}{2\pi} \int d^2 \mathbf{k} \nabla \times i \langle u(\mathbf{k}) | \nabla_{\mathbf{k}} | u(\mathbf{k}) \rangle,
\end{equation}
where $u(\mathbf{k})$ is the periodic part of the Bloch wave function of the lower occupied band. For the $M<0$ region, the calculated Chern number, $C=0$, corresponding to a BI. The Chern number $C=-1$ in the $M>0$ region gives rise to a QAH state. In contrast to the $B<0$ case, we have also checked the band topology for $B>0$ by calculating the Chern number, where $M>0$ and $M<0$ lead to BI and QAH state, respectively. That is, the band topology of the system becomes determined by  the sign of $MB$ instead of $mB$. If $MB<0$, the system is a QAH state, while a magnetic band insulator appears with $MB>0$.

Combining the above calculation of the topological invariant and the corresponding phase diagram, it is seen that for an initial BI, a topological phase transition to the QAH state occurs with the increase of the the attractive interaction; but for the exsiting QAH state, there is no topological phase transition existing no matter how the attractive interaction changes. It is well understood by the definition of $\Delta(U)$ associated with the occupation number expectation. To be specific, we begin with a BI in the absence of the fermion interaction, where $m$ has a negative sign for the adopted negative $B$ ($mB>0$). Hamiltonian (\ref{little_hp}) indicates that the fermions prefer to occupy the spin-up states in the whole Brillouin zone, i.e., $\langle n_{i\uparrow} \rangle >>\langle n_{i\downarrow} \rangle$. The occupation priority keeps unchanged with the increase of $U$. For the half-filling case, one can obtain $\langle n_{i\uparrow} \rangle > \frac{1}{2}$. According to the definition $\Delta(U) = U \left( \frac{1}{2} - \langle n_{i\uparrow} \rangle \right)$, $\Delta(U)$ has a minus sign which is the same with that of $m$. Therefore, $M$ ($M=m-\Delta(U)$) might experience a sign change with the increase of $U$, leading to a topological phase transformation to topological QAH state.
On the other hand, if the initial state is a QAH state with positive $m$ and negative $B$ ($mB<0$), the fermions prefer to occupy the spin-up states when the $k$ is far away from the center point, leading to $\langle n_{i\uparrow} \rangle > \langle n_{i\downarrow} \rangle$. We have $\langle n_{i\uparrow} \rangle > \frac{1}{2}$ for the half-filling case and $\Delta(U) = U \left( \frac{1}{2} - \langle n_{i\uparrow} \rangle \right)<0$. Since $\Delta(U)$ and $m$ have the opposite signs, $M$ has the same sign with $m$, which means the topology of the system keeps unchanged though the attractive interaction is introduced.

\subsection{\label{sec4}The enlarged topological energy gap}
\begin{figure}
\includegraphics[width=8.5cm]{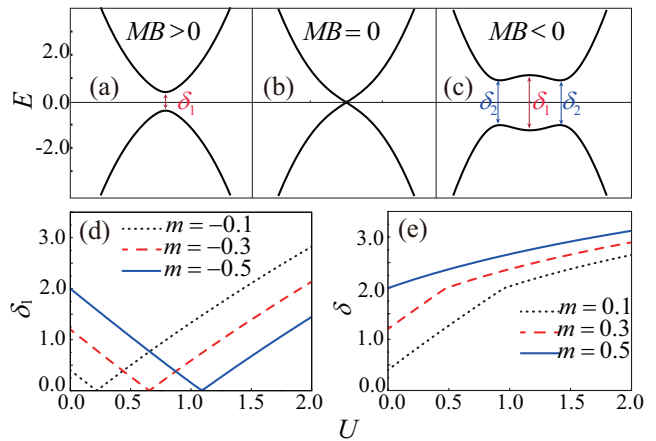} \caption{(Color online) The band structures when (a) $MB>0$, (b) $MB=0$, (c) $MB<0$, where $\delta_1$ represents the energy gap at the $k=0$ point and $\delta_2$ represents the gaps at the anti-crossing points. (d) and (e) The evolution of $\delta_1$ and $\delta$ with the increased $U$ for $m<0$ and $m>0$, respectively. The parameter $B$ is set as $B=-1$.}
\end{figure}
We further study the evolution of the band gap under the action of the attractive interaction. Based on the continuum model with the renormalized mass $M$, the dispersion relation is obtained as
\begin{equation}
E=\pm\sqrt{Ak^2+(M+Bk^2)^2},
\end{equation}
with the band gap $\delta_1=2|M|$ at the $k=0$ point.

Figures 3(a-c) show the energy spectra when $M=-0.4~(MB>0)$, $M=0~(MB=0)$ and $M=1.2~(MB<0)$, respectively. It is seen that the band gap at the $k=0$ point, $\delta_1$, undergoes an opened-closed-reopened process when the system is transformed from the BI state to the QAH state. Figure 3(d) further quantitatively shows the evolution of $\delta_1$ with the increased attractive interaction $U$, which starts with an initial BI with $m<0$. $\delta_1$ first declines to zero and then rises linearly with the increase of $U$, agreeing with the Fig.~3(a-c) and confirming a band inversion process with the variation of $U$.

Besides $\delta_1$ at $k=0$, there are another two local minimum band gaps, $\delta_2$, for the QAH state, which are opened by the spin-orbit coupling and localized at the two crossing points of the two subbands, as mentioned in Fig.~1(c) and shown in Fig.~3(c). $\delta_2$ is computed as
\begin{equation}
\delta_2=2|A|\sqrt{-\frac{M}{B}}.
\end{equation}
Therefore, the global band gap, $\delta$, of the QAH state is determined by the minimum between $\delta_1$ and $\delta_2$, that is, $\delta=\text{Min}(\delta_1,\delta_2)$. We have $\delta=\delta_1$ for $|MB|<A^2$ and  $\delta=\delta_2$ for $|MB|>A^2$. The evolution of the global gap of the QAH state  with $U$ is plotted in Fig.~3(e). It is seen that $\delta$ increases with two distinct slopes in two region of $U$, corresponding to two above local band gaps. Moreover, the attractive interaction can effectively enlarge the system gap of the QAH state, which is of importance for realizing the robust topological insulating properties within a large energy window, given that most of two-dimensional topological insulating states in the experiments have small band gaps of only several tens of meV.

\subsection{\label{sec4}The effect of the attractive interaction on the edge states}

We then investigate the effect of the attractive Hubbard interaction on the edge state of the QAH nanoribbon based on the square lattice model, as shown in the schematic inset of Fig.~4. For an infinite-wide nanoribbon in the QAH state, the left-propagating and right-propagating edge states are well localized at two sides of the nanoribbon, respectively. The edge states should be decoupled without the inter-edge overlap of corresponding wavefunctions. However, for a nanoribbon with a finite wide, the inter-edge crosstalk is unavoidable due the finite-size effect\upcite{WLiu, LFMiao} and need to be weakened in order to keep the edge state dissipationless even in the low-energy regime.

The band structure of the QAH nanoribbon without and with attractive Hubbard interaction are computed, as shown in Fig.~4(a) and 4(b), where we use a nanoribbon that has 20 lattice sites along the transverse direction.
When the interaction is absent, the bulk band gap is about $0.5$. In a large energy range, the counterpropagating edge states keep good linearity within the band gap [Fig.~4(a)]. Zooming in on the low-energy regime, the edge states are actually gapped at $k=0$, with a small value of about $5\times10^{-3}$ [the inset of Fig.~4(c)], due to the edge-state coupling between two edges. When the attractive interaction is introduced, the bulk band gap is enlarged up to $1.2$ and the gap of the edge states become negligible, compared with $U=0$. In the amplified view, there is no edge gap but a perfect linear dispersion, as shown in the inset of Fig.~4(d).

For the certain width, the smaller edge gap of the nanoribbon arises from the more localized edge wavefunction, which are comfirmed by calculating the edge state probability density $|\psi|^2$. Figures 4(c-d) show the real-space probability distribution of the edge states with the nearly zero energy. For $U=0$, the probability densities of the two edge states peak at the second outmost sites and they gently decay when moving towards the middle of the nanoribbon, as shown in Fig.~4(c). There is a small overlap between the tails of the edge probability densities, which leads to a gap opening of edge states. When the attractive interaction is added, the probability densities of the two edge states in Fig.~4(d) peak at the outmost sites and drop rapidly to zero, demonstrating a more localized probability distribution compared with the one in Fig.~4(c). Since the more localized edge states reduce the inter-edge overlap between edge wavefunctions, the edge states show a perfect linear dispersive in the observable energy scale, which provides a better platform for dissipationless quantum transport even at the low-energy limit. It is also noted that the more localized edge states and their perfect linear dispersion can be regarded as a result of enlarged bulk energy gap of the QAH state under the action of the attractive interaction, since the localization degree of the edge states is proportional to the magnitude of the bulk energy gap.\upcite{Volkov, Dolcetto}

\begin{figure}
\includegraphics[width=8cm]{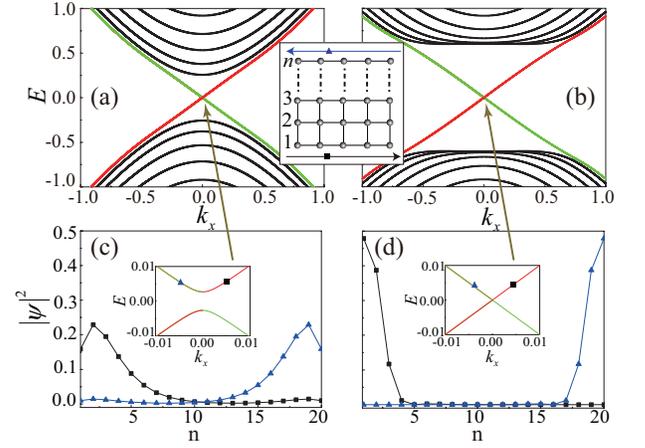} \caption{(Color online) The band structures of the QAH nanoribbon based on the square lattice model when (a) $U=0$ and (b) $U=1$. The edge state probability density, $|\psi|^2$, of the nanoribbon for (c) $U=0$ and (d) $U=1$, where the triangles and squares respectively correspond to the left-propagating and right-propagating edge states near the zero energy. The schematic of the nanoribbons used in the lattice model is shown in the inset of (a) and (b), while the insets of (c) and (d) zoom in on the low-energy states. The width of the ribbon, $n$, is set to 20. The parameters $B=-1$ and $m=0.2$.}
\end{figure}

\section{\label{sec5}SUMMARY}
To conclude, we have investigated the effect of the on-site attractive interaction on topological properties of the two-band magnetic Dirac fermion model based on a square lattice within the mean-field approximation. By the self-consistent calculation, it is found that a topological phase transition occurs from the BI state to the QAH state, with the increase of the attractive interaction. For an exsiting QAH state, there is no topological phase transition, but the topological energy gap can be effectively enlarged under the action of the attractive interaction. Moreover, the attractive interaction restores the linear dispersion of the edge state even at the low-energy limit, since more localized edge states reduce the inter-edge coupling. A large bulk gap and perfect linear edge dispersion will help improve the performance of QAH insulator and realize energy-efficient edge quantum transport.

\section{\label{sec6}Acknowledgements}

This work was supported by the National Natural Science Foundation of China (Grant No.~11504084, 11647164, 11890703 and 11805103), and the Natural Science Foundation of the Jiangsu Higher Education Institutions of China (Grant No.~18KJB140005, 17KJD170004 and 16KJB140008).

\end{document}